# Intelligent Agent Based Semantic Web in Cloud Computing Environment

Debajyoti Mukhopadhyay[1], Manoj Sharma[1], Gajanan Joshi[1],
Trupti Pagare[1], Adarsha Palwe[1]

Department of Information Technnology[1]
Maharashtra Institute of Technology
Pune 411038, India
{debajyoti.mukhopadhyay, manojsharma2708, joshi.gajanan09,
pagaretrupti2, adarsha125}@gmail.com

**Abstract:** Considering today's web scenario, there is a need of effective and meaningful search over the web which is provided by Semantic Web. Existing search engines are keyword based. They are vulnerable in answering intelligent queries from the user due to the dependence of their results on information available in web pages. While semantic search engines provides efficient and relevant results as the semantic web is an extension of the current web in which information is given well defined meaning. MetaCrawler is a search tool that uses several existing search engines and provides combined results by using their own page ranking algorithm. This paper proposes development of a meta-semantic-search engine called SemanTelli which works within cloud. SemanTelli fetches results from different semantic search engines such as Hakia, DuckDuckGo, SenseBot with the help of intelligent agents that eliminate the limitations of existing search engines.

## 1. Introduction

Search Engines has become one of the most important and used tool over the World Wide Web. Search Engines normally search web pages for the required information and then display the results by using ranking algorithms. These commercially available search engines do not completely serve the needs and demands of the users. Most of the time, the relevance of the provided results are not accurate. These types of problem occur because of the structure of the current Web.

### 1.1 Intelligent Agent

Intelligent Agents are an emerging technology that is making computer systems easier to use by allowing people to delegate work back to the computer. They help to do things like finding and filtering information, customize views of information and automate work. An intelligent agent is software that assists people and acts on their behalf. Agents can intelligently summarize data, learn from us and even can make recommendations to us.

- Delegation aspect of intelligent agents, which is certainly one thing that sets agents apart, since agents are built to help people, just like human assistants.
- All agents are autonomous that is an agent has control over its own actions.
- All agents are goal driven. Agents have a purpose and act in accordance with that purpose.

**1.2 Cloud Computing**

Cloud Computing is a model in which customers plug into the cloud to access IT resources which are priced and provided on demand. IT resources are rented and shared among multiple tenants much as office space, apartments or storage spaces used by tenants. Delivered over an Internet connection, the cloud replaces the data company center or server providing the same service.

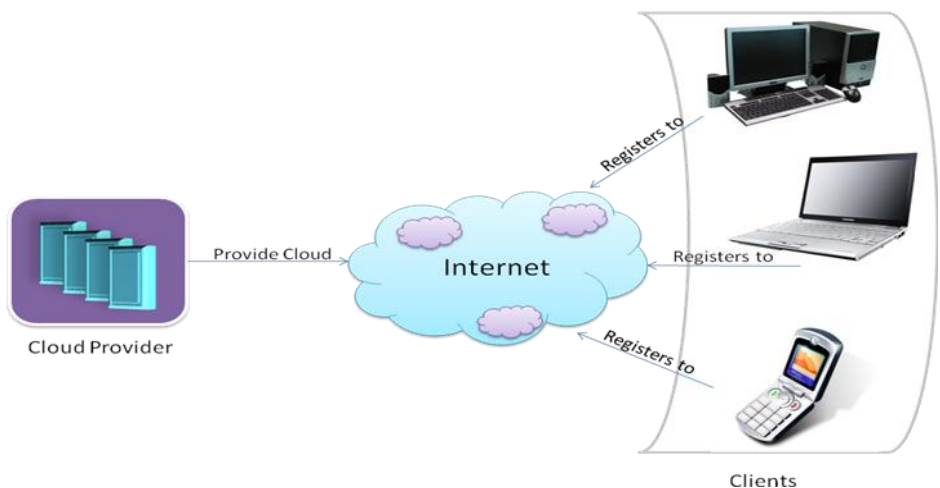

**Fig. 1.** Simple Cloud

Cloud makes it possible for us to access our information from anywhere at any time. At any time it removes the need for us to be in the same physical location due to the following features:

- **SaaS (Software as a Service)**
  It provides all the functions of a sophisticated traditional application to many customers and often thousands of users, but through a Web browser, not a locally-installed application.
- **PaaS(Platform as a Service)**
  Delivers virtualized servers on which customers can run existing applications or develop new ones without having worry about maintaining operating systems, server hardware, load balancing or computing capacity.

- **IaaS(Infrastructure as a Service)**
  Delivers utility computing capability, typically as raw virtual servers, on demand that customers configure and manage.

**1.3 Semantic Search Engine**
Here we describe in brief, the semantic search engines used in SemanTelli:

- **DuckDuckGo**

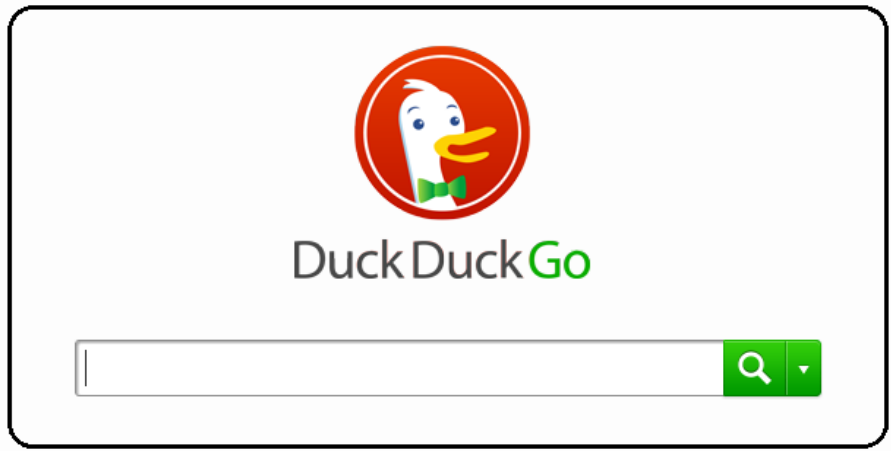

**Fig. 2.** DuckDuckGo semantic search engine

The main sources of information for DuckDuckGo are Crowd-sourced websites that makes it capable to enhance traditional results and improves their relevance. It provides policy that values privacy and does not record user information as users are not profiled.

- **Hakia**

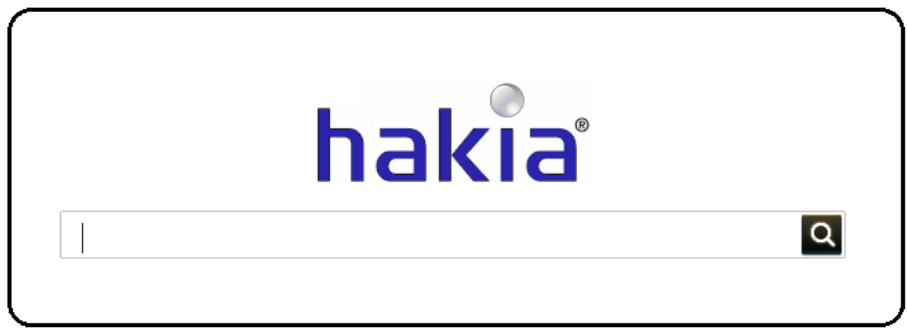

**Fig. 3.** Hakia semantic search engine

Hakia uses enhanced semantic technology for ranking algorithm that uses QDEX (Query Detection and Extraction). It perceives quality results from all segments like news, blogs, credible, hakia galleries, images and videos. For long queries, it highlights relevant phrases, keywords or sentences that relates to users query.

- **Sensebot**

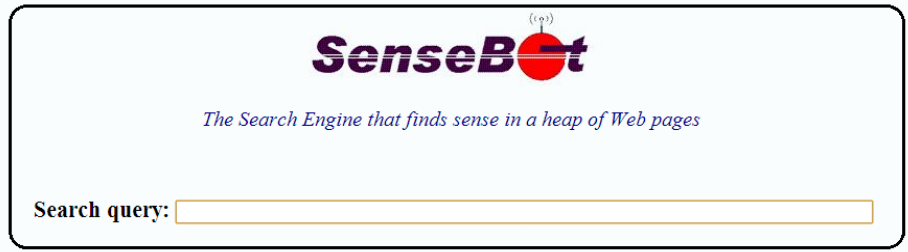

**Fig. 4.** Sensebot semantic search engine

Sensebot prepares the text summary according to the user's search query. It identifies key semantic concepts by using text mining algorithms that parse the Web Pages. The retrieved multiple documents are then used to perform a coherent summary. This coherent summary becomes the final result for user's query. The main sources for these results are usually the news agencies.

## 2. SemanTelli Architecture

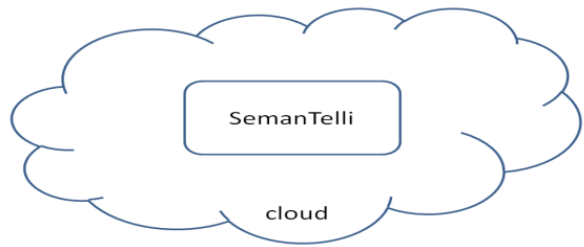

**Fig. 5.** SemanTelli within Cloud

SemanTelli is a simple meta-semantic-search engine which extends the concept of meta-search engines to enhance searching of specific data within cloud. Figure 5 provides the visualization of this concept. Figure 6 shows the whole architecture of SemanTelli with its major functioning blocks.

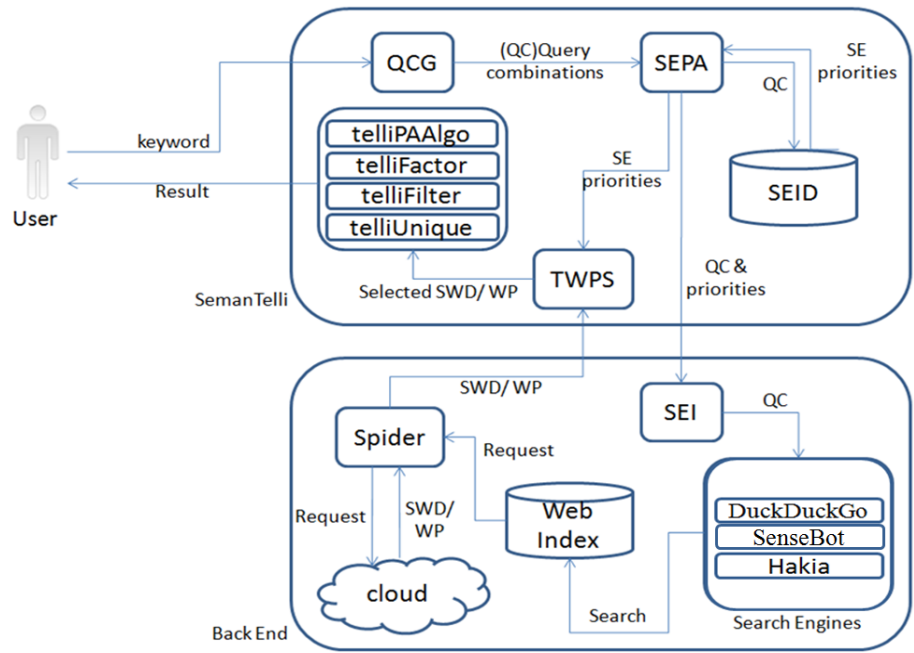

**Fig. 6.** Architecture of SemanTelli

Abbreviations used:
        QCG  : Query Combination Generator
        SEPA  : Search Engine Priority Assigner
        SEI    : Search Engine Interactor
        TWPS : Temporary Web Page Storage
        SEID  : Search Engine Information Database

Our proposed architecture is comprised of two blocks, as given below:

**2.1 SemanTelli**
SemanTelli comprises of user interface and other modules to find out search engines (SE) priorities and to filter the obtained search results from search engines. Query Combination Generator (QCG) accepts the query from user and generates its possible and meaningful combinations. These combinations are sent to SEPA (Search Engine Priority Assigner) which finds the priority for different SEs to monitor search results. SEPA uses Search Engines Information Database (SEID) which contains information about importance of each search engine for particular keyword or its domain. This information helps SEPA to give specific priorities to each SE. SEPA sends these priorities and QCs to SE interface (SEI). The search results consisting of Semantic Web Documents (SWD) and Web Pages (WP) are returned by backend to Temporary Web Page Storage (TWPS), which stores the results temporary. TEMPS uses the properties from SEPA to filter out the SWD/WPs and sends the required results for further refinements.

Here, many selected SWD/WPs have to go through our algorithm in order to reduce redundancy and provide actual needed output to the user. Our approach uses a parameter named "telliFactor" to decide arrangement position of each SWD. telliFactor is calculated using many criteria such as frequency of QC, number of links and its hit count. After arranging their properly in decreasing order of telliFactor, final output is shown to the user.

**2.2 Back End**
This block gets the QC and SE priorities from SEPA and searches the SWDs and WPs over the World Wide Web through our different semantic search engines. Here, spiders located over the cloud helps search the web pages on various clouds. Spiders also send these web pages to TWPS for further processing and refinement.

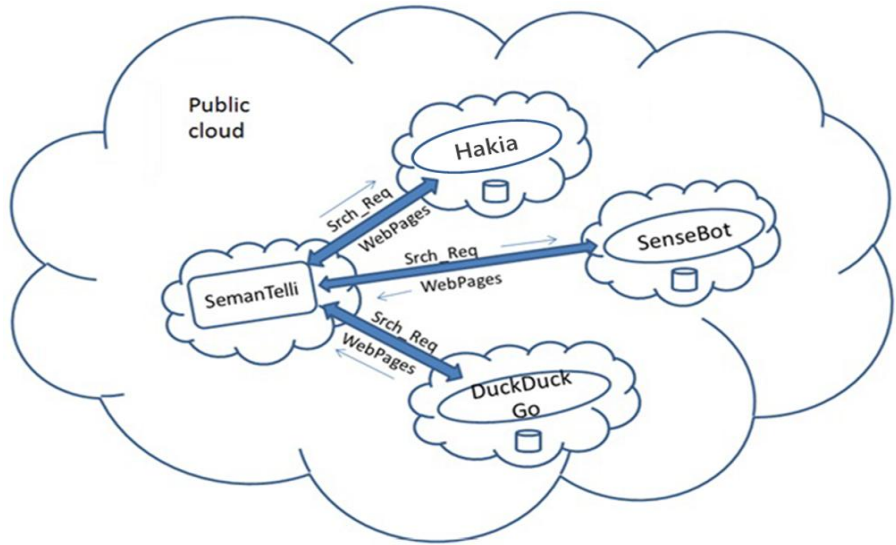

**Fig. 7.** SemanTelli in cloud environment

Figure 7 throws light on the whole concept of SemanTelli and its working on multiple cloud environment. SemanTelli will send the search requests to multiple search engines that existed in different clouds. This search request would make the SEs to return Web Pages satisfying that query. Cloud Page Indexes located in each cloud provides fast searching of these web pages as Spider does in WWW.

## 3. Current status of SemanTelli

SemanTelli is in developing phase. It fetches the result from semantic search engines, finds telliFactor($t_F$) for each result and arranges these results. We have assigned initial weight to each fetched result as mentioned in Table. 1.

| Sr. No. | Search Engine | Initial Weight ($W_i$) |
|---|---|---|
| 1 | DuckDuckGo | 0.3 |
| 2 | Hakia | 0.2 |
| 3 | SenseBot | 0.1 |

Table 1. Initial Weight

Considering relevance of results, we have assigned more initial weight to DuckDuckGo results as compared to Hakia and SenseBot results.

Figure 8 highlights the implementation details of SemanTelli. It shows that fetched results are temporarily stored in respective buffers. These results are traversed in order to compute telliFactor which is our relevance score. After post processing, results are arranged with respect to telliFactor.

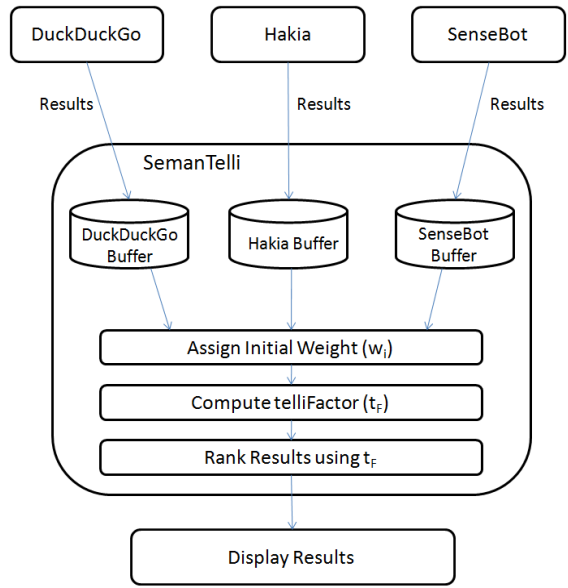

**Fig. 8.** Implementation details

Following algorithm is used to compute the telliFactor:

```
Algorithm: telliFactor
Input:     Obtained results
Output:    telliFactor of results

Consider damping factor d = 0.85.
Step1:  Select redundant result and increment its weight
        by 0.1.
Step2: Calculate relevance factor (r_F) for each Web page.
```

$$r_F = \frac{(h + 1)}{1000}$$

```
        where h = hit count of Web page
              l = number of out-links
Step3:  For each result
            Tellifactor(t_F)  =   ( W_i  *  d )  +   r_F
```

The results are arranged according to the telliFactor and if the telliFactor of two or more results is identical then their rank from respective search engine is considered.

Fig 9 shows home page of SemanTelli.

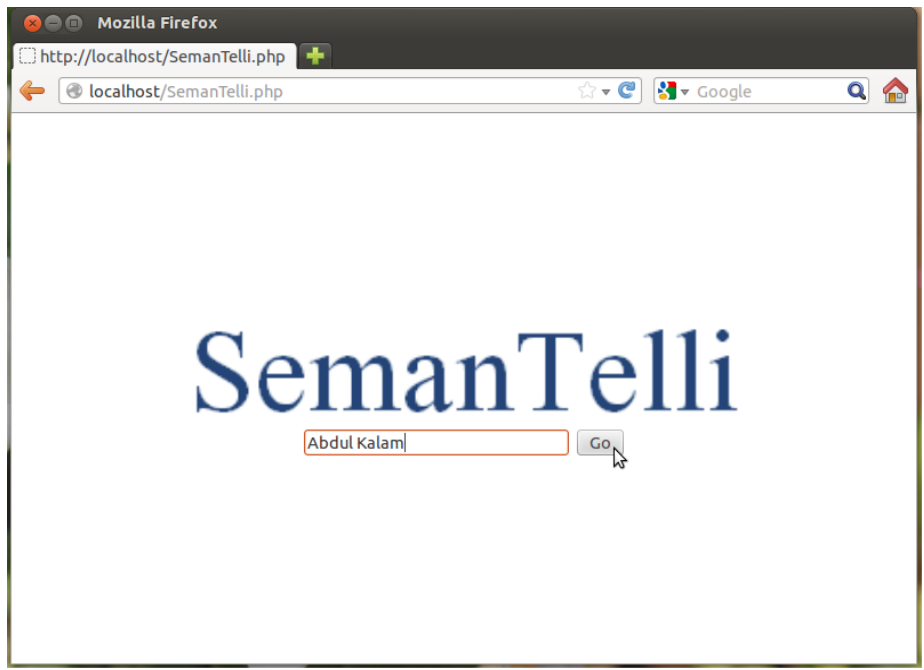

**Fig. 9**. SemanTelli Interface

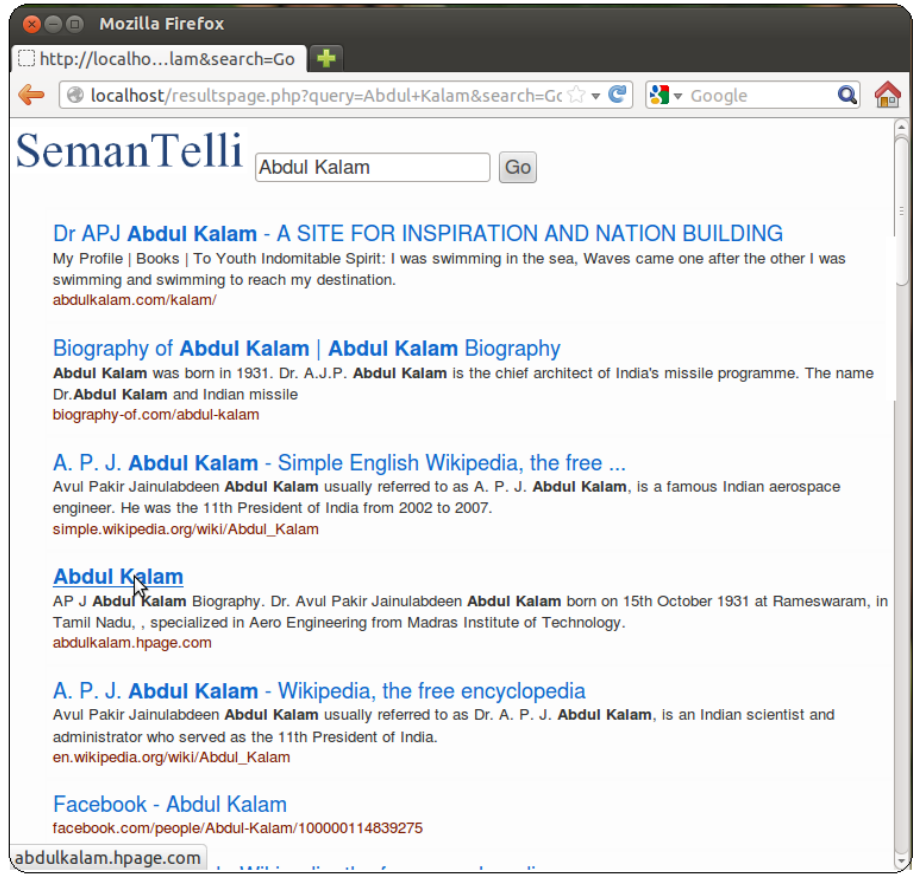

**Fig. 10**. SemanTelli query results

## 4. Conclusion

We developed a meta-semantic-search engine "SemanTelli" which highlights the novel concept of integrating search results from existing semantic search engines such as DuckDuckGo, Hakia, SenseBot within cloud computing environment. This will provide users with an effective way of finding out required results from many clouds. SemanTelli mines deeply through the combinations of user query using semantic search engines and brings out only relevant results from the huge information present in cloud environment.